\documentclass[%
reprint,
 amssymb, amsmath,%
 aps,
 prl,
]{revtex4-1}

\usepackage[pdftex]{graphicx}
\graphicspath{{./../figures/}}
\DeclareGraphicsExtensions{.pdf,.jpeg,.png}

\usepackage{stmaryrd}
\usepackage{docs}%
\usepackage{bm}%
\usepackage[colorlinks=true,linkcolor=blue]{hyperref}%
\expandafter\ifx\csname package@font\endcsname\relax\else
 \expandafter\expandafter
 \expandafter\usepackage
 \expandafter\expandafter
 \expandafter{\csname package@font\endcsname}%
\fi
\begin{document}

\title{Tests of Sapphire Crystals Produced with Different Growth Processes for Ultra-stable Microwave Oscillators.}
\author{Vincent Giordano$^{1}$}%
\email{giordano@femto-st.fr}
\author{Christophe Fluhr$^{1}$, Serge Grop$^{1}$, Beno\^{i}t Dubois$^{2}$.}

\affiliation{\vskip 5mm $^{1}$ FEMTO-ST Institute - UMR 6174\\
CNRS / ENSMM / UFC / UTBM\\
26 Chemin de l'\'Epitaphe\\
25000 Besan\c{c}on -- FRANCE}%

\affiliation{\vskip 5mm $^{2}$ FEMTO Engineering\\
32 avenue de l'Observatoire \\
25000 Besan\c{c}on -- FRANCE\\}%
\vskip 5mm
\date{2015 April, 10}%


\begin{abstract}
We present the characterization of  $8-12$ GHz whispering gallery mode resonators machined in high-quality sapphire crystals elaborated with different growth techniques. These microwave resonators are inttended to constitute the reference frequency  of ultra-stable Cryogenic Sapphire Oscillators. We conducted systematic tests near 4K on these crystals to determine the unloaded Q-factor and the turnover temperature for whispering gallery modes in the 8-12 GHz frequency range. These characterizations show that high quality sapphire crystals elaborated with the Heat Exchange or the Kyropoulos growth technique are both suitable to meet a fractional frequency stability better than $1\times 10^{-15}$ for 1 s to 10.000 s integration times. 
\end{abstract}

\maketitle

\section{Introduction}

State-of-the-art ultra-stable microwave oscillators are currently based on a whispering gallery mode sapphire resonator operated in the range $8-12$ GHz. The best near carrier phase noise is achieved with a room temperature sapphire resonator oscillator incorporating a sophisticated electronic degenerating the noise of the sustaining oscillator stage \cite{ivanov98}. On the other hand, the Cryogenic Sapphire Oscillator (CSO) in which the sapphire resonator is cooled near 6 K, provides a fractional frequency stability better than  $1\times 10^{-15}$ for integration times lower than 10,000 s \cite{eftf-2012-cso,hartnett-2012-apl}. 
The recent demonstration of a low maintenance CSO based on a pulse-tube cryocooler paves the way for its deployment in real field applications \cite{RSI-2012}.  Such oscillators are required for the operation of laser cooled atomic clocks \cite{santarelli99-prl,ifcs-2015-fountain}. The CSO can provide the means to improve the resolution of the space vehicules ranging and Doppler tracking provided by Deep Space Networks as well as those of Very Long Baseline Interferometry (VLBI) Observatories  \cite{rsi10-elisa,ell11-elisa} . The CSO can also enhance the calibration capability of Metrological Institutes or help the qualification of high performances clocks or oscillators \cite{AppPhysB-2014}. \\

The CSO exceptional performances result from the intrinsic properties of the material in which the resonator is machined. A high Q-factor requires a monocrystal with a low structural defects density. With a high quality sapphire crystal at 4 K, Q-factor of  $1\times10^{9}$ are typically obtained for whispering gallery modes around 10 GHz. Nevertheless the sapphire crystal should not be a perfect one. Indeed the thermal compensation induced by the accidental paramagnetic impurities that substitute to Al$^{+3}$ is essential for the achievement of the highest frequency stability. The main issue is that the impurities concentration giving a suitable turnover temperature between 5-8 K is very low (typically less than 1 ppm) and can not be warranted by the manufacturer. To build a CSO the designer has no other choice than to buy expensive crystals without knowing if they are suitable for his application. A cryogenic test is the only way to reveal the temperature dependance of the resonator frequency and thus to fix the operating temperature.\\

In this paper we present the characterization of cryogenic sapphire resonators produced with different growth techniques and by different manufacturers. Although far to be exhaustive, the results presented in this paper will help to the building of an optimized CSO and show that state-of-the-art frequency stability can be obtain with several kinds of sapphire crystal.

\section{The quest for a $1\times 10^{-15}$ frequency stability}
The CSO incorporates as a frequency reference a cylindrical sapphire resonator placed in the center of a copper cavity that can be cooled down to 4 K. In this structure, high order whispering gallery modes are characterized by a high energy confinement in the sapphire disk due the total reflection at the vacuum-dielectric interface. The different resonators we designed operate on quasi-transverse magnetic whispering galery modes as $WGH_{m,0,0}$ where $m$ is the number of wavelength in the resonator along the azimuth. To operate in the $8-12$ GHz frequency range, the resonator diameter is $\Phi=30$-$50$ mm and its  thickness $H=20$-$30$ mm. If the azimuthal index is sufficiently high  ($m \geq 13$ typically) the Q factor can achieve $1\times10^{9}$  at the liquid-He temperature.\\

The CSO is Pound-Galani oscillator \cite{galani84}. In short, the resonator is used in transmission mode in a regular oscillator loop, and in reflexion mode as the discriminator of the classical Pound servo \cite{pound46}. In an autonomous cryocooled CSO, the sapphire resonator is placed into a cryostat and in thermal contact with the second stage of a pulse-tube cryocooler delivering typically 0.5 W of cooling power at 4 K.  The resonator temperature can be stabilized above 4 K at $\pm 200~\mu$K. Details on the design and on the techniques that have to be applied to get an ultimate frequency stability can be found in \cite{locke08,rsi10-elisa,uffc-2011-long-term-stab,hartnett10-uffc}. \\

Our goal is to provide an ultra-stable oscillator that meets the most stringent short term frequency stability specifications, such as those for the Deep Space Network for satellites and space vehicles navigation. In this way, let us specify a short-term fractional frequency stability expressed in term of Allan standard deviation as:

\begin{equation}
\sigma_{y}(\tau) \leq 1\times 10^{-15} \mathrm{~for~ 1~s}\leq \tau \leq 10,000 \mathrm{~s}
\end{equation}

The question we address is what are the required resonator properties  and which type of sapphire material is suitable to meet this specification.

\subsection{Resonator design and adjustment}
As explained in \cite{rsi10-elisa}, in our most advanced CSO, a $\Phi=54$ mm, $H=30$ mm resonator is operated on the $WGH_{15,0,0}$ mode near 10 GHz.
The input coupling coefficient $\beta_{1}$ has to be set near unity to optimize the Pound discriminator. The ouput coupling coefficient $\beta_{2}$ should be low enough to not degrade the loaded Q-factor but to stay compatible with the gain of the sustaining amplifier. Overall insertion losses through the cryostat of $-30$ dB or so is generally a good trade-off. With optimized coupling coefficients, i.e. $\beta_{1} =1$ and $\beta_{2} \ll1$, the loaded Q-factor is the half of the unloaded one: $Q_{L} \approx \frac{1}{2} Q_{0}$.

\subsection{Short term frequency instability: The Line Splitting Factor}
Noise in the electronics is the main source of short-term instabilities; how exactly this noise affects the frequency stability depends on the oscillator configuration and on $Q_{L}$. Without characterizing the exact nature of the oscillator's individual noise sources, one would like a rough estimate of what the resonator's bandwidth should be to achieve a given frequency stability. It is generally admitted that the attainable frequency stability $\sigma_y(\tau)$ at a given time interval (e.g.~$\tau=1$ s) cannot be lower than a given fraction of the resonator bandwidth $\Delta \nu$\cite{tobar00}. We thus define an empirical figure of merit, namely the "Line Splitting Factor", or $LSF$, as the ratio of the frequency fluctuations $\delta \nu$ averaged over $\tau$ of the generated signal divided by the resonator's bandwidth. If $\nu$ is the signal frequency, the $LSF$ is:
\begin{equation}
LSF(1\mathrm{s})=\frac{\delta\nu}{\Delta\nu}=Q_{L} \times \dfrac{\delta\nu}{\nu}=Q_{L} \times \sigma_{y}(1s)
\end{equation}

The $LSF$ provides a way of quantifying the overall effect of the noise associated with the oscillator's
electronics. Considering the best experimental results obtained to date \cite{hartnett-2012-apl, locke08,jap-2014}, we assume $LSF \approx 2.5\times 10^{-7}$, which is a conservative value. Thus, to get a stability of  $1 \times 10^{-15}$, the minimum
resonator unloaded Q-factor is:
\begin{equation}
Q_{0}=2\times Q_{L}=2~\dfrac{LSF(1\mathrm{s})}{\sigma_{y}(1\mathrm{s})} \geq 500 \times 10^{6}
\label{equ:Q-factor}
\end{equation}

\subsection{Medium term frequency limitation}

The resonator frequency is determined by its geometry and by the wave velocity inside the medium.
These physical characteristics are in turn affected by the resonator's temperature and the power of the injected signal. 
The resonator's sensitivity to these environmental parameters limits the oscillator's medium term frequency stability. \\

\subsubsection{Thermal sensitivity}

For an ideal resonator made in pure sapphire, although the frequency thermal sensibility would decrease significantly with temperature, it will never be low enough to achieve the target stability with a state-of-the-art temperature controler. Fortunately, it turns out that high-purity sapphire crystals always contain a small
concentration of paramagnetic impurities, as Cr$^{3+}$, Fe$^{3+}$ or Mo$^{3+}$. At low temperatures, these residual impurities induce a magnetic susceptibility $\chi$ whose temperature dependence can in certain circumstances compensate the intrinsic sensitivity of the pure sapphire. Following the notation introduced by Mann \cite{mann92-jpd}, for a high-order whispering gallery mode  the thermal dependance of the resonator frequency $\nu$ can be written as:\\
\begin{equation}
\dfrac{\nu(T)- \nu_{0K}}{\nu_{0K}} = A T^{4}+ \frac{1}{2}~\chi'(\nu,T)
\label{equ:f-T-sens}
\end{equation}  
where, $\nu_{0K}$ would be the mode frequency at 0K in a pure sapphire resonator. $A\approx -3\times 10^{-12}$ K$^{-4}$ combines the temperature dependance of the dielectric constant and the thermal expansion \cite{luiten96}. 
$\chi'$ is the real part of the susceptibility induced by the paramagnetic dopants. If the sapphire crystal contains a density $N$ of a paramagnetic ion characterized by its Electron Spin Resonance (ESR) frequency $\nu_{j}$ and a spin-to-spin relaxation time $\tau_{2}$, $\chi'(\nu)$  is a dispersive lonrentzian function that nulls at $\nu_{j}$ \cite{siegman_maser}:

\begin{equation}
\chi'(\nu,T)= \chi_{0}(T) \dfrac{(2\pi\tau_{2})^{2}(\nu-\nu_{j}) \nu_{j}}{1+(2\pi\tau_{2})^{2}(\nu-\nu_{j})^{2}}
\label{equ:chiprim}
\end{equation}
The thermal dependance is contained in the dc-susceptibility $\chi_{0}(T)$, which results from the distribution of the ions on their energy levels through the effect of the thermal agitation. Assuming the ion in the crystal lattice behaves like a free spin $S$, $\chi_{0}(T)$ is expected to follow the Curie law \cite{siegman_maser}:

\begin{equation}
\chi_{0}(T) = N   \dfrac{\mu_{0}g^{2} \mu_{B}^{2}  }{3 k_{B}T} S(S+1)
\label{equ:chi0C}
\end{equation}
where $\mu_{0}$ is the permeability of free space and $k_{B}$ the Boltzmann constant.
To make explicit the $1/T$ dependance of the susceptibility, let us introduce $C(\nu)$ as:
\begin{equation}
\frac{C(\nu)}{T}= \frac{1}{2}~\chi'(\nu,T) 
\end{equation}
The equation \ref{equ:f-T-sens} can be rewritte as :
\begin{equation}
\dfrac{\nu(T)- \nu_{0K}}{\nu_{0K}} = A T^{4}+ \frac{C(\nu)}{T}
\label{equ:f-T-sens-2}
\end{equation}

A thermal compensation can occur if the derivative of equation \ref{equ:f-T-sens-2} nulls, which imposes  $C(\nu)<0$, and thus $\chi'(\nu,T)<0$, which means $\nu <\nu_{j}$, i.e the signal frequency is below the ESR. In this case, the temperature $T_{0}$ at which the resonator thermal sensitivity nulls is: 
\begin{equation}
T_{0}=\left ( \dfrac{C(\nu)}{4A} \right )^{1/5}
\label{equ:T0}
\end{equation}

The ESR characteristics of ions commonly found in sapphire is given in the table \ref{tab:ions}. We reported also for each paramagnetic specy the value of $T_{0}$ calculated from the equation \ref{equ:T0} for $N=1$ ppm and at 10 GHz.

\begin{table}[h!!!!!]
\centering
\caption{ESR of the Iron-group paramagnetic ions that can be found in high-purity sapphire crystals.}
\label{tab:ions}
\begin{tabular}{lcccc}
\hline
Ion	&$S$	&$\nu_{j}(GHz)$	&$\tau_{2}$ (ns)	&$T_{0}(K)$ \\
	&		&        			&				&$1$ppm/10 GHz			\\	
\hline
\hline
Cr$^{3+}$  	&3/2		& 11.4~~	 	& 7	&		12.6	\\
Fe$^{3+}$  	& 5/2	& 12.0~~ 		& 20 &		 13.9\\
Mo$^{3+}$  	& 3/2	& 165.0~~		& 12	&		 8.4\\
\hline
\end{tabular}
\end{table} 

\begin{figure}[ht!]
	\centering
	\includegraphics[width=\columnwidth]{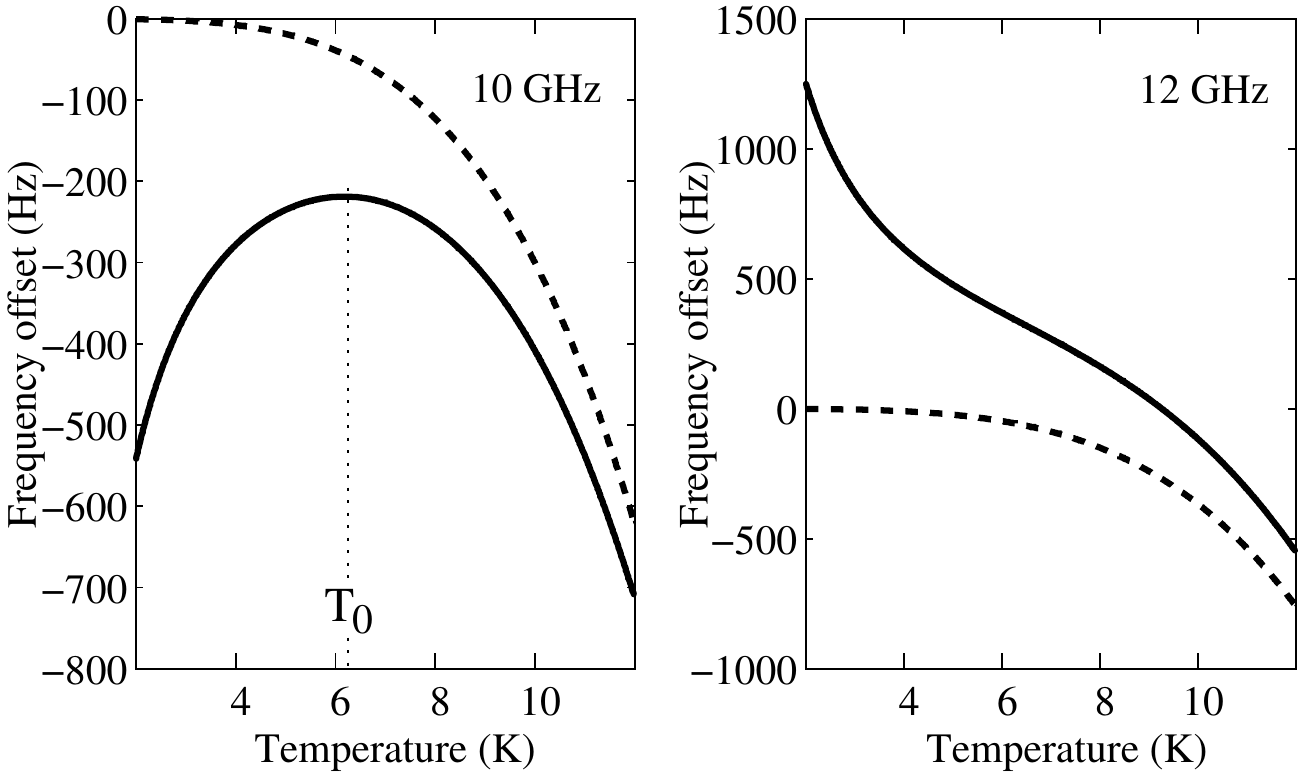}
	\caption{Calculated frequency-to-temperature evolution of the $WGH_{15,0,0}$(10 GHz) and $WGH_{19,0,0}$ (12 GHz) of a $54\times30$ mm sapphire resonator. Dashed lines: pure sapphire crystal, Bold lines: sapphire crystal containing 10 ppb in weight of Cr$^{3+}$}.
	\label{fig:figure-1}
	\end{figure}

It is clear that an impurity density such as $N=1$ ppm is too large to get a turnover point below 8 K, which is required to optimize the cryogenerator operation and to keep a high Q-factor. $T_{0}$ being propotional to $N^{1/5}$, the Cr$^{3+}$ and Fe$^{3+}$ concentrations should be very low as their ESR frequency is near 10 GHz. The figure \ref{fig:figure-1} show the behavior of two modes situated at each side of the Cr$^{3+}$ ESR. These frequency-to-temperature evolutions have been calculated using the previous equations and assuming a Cr$^{3+}$ concentration of 10 ppb only.

The $WGH_{15,0,0}$ mode at 10 GHz shows a turnover temperature $T_{0}\approx 6$K. Around  $T_{0}$  the mode frequency can be approximated by a quadratic function of the temperature. Stabilized at $T_{0}$ the resonator frequency is no longer sensitive to the temperature fluctuations and a high frequency stability can be achieved. Conversely the frequency of the mode $WGH_{19,0,0}$ is above the Cr$^{3+}$ ESR frequency and no turnover can be observed making it no suitable to build an ultra-stable oscillator.
%

\subsubsection{Power sensitivity}
The CSO sensitivity to the injected power results mainly from the thermal effect and from the radiation pressure, which becomes noticeable when the Q-factor is high. At a signal power higher than typically 100 $\mu$W, the resonator sensitivity is of the order of $1\times 10^{-8/}$W. At a lower power this sensitivity is affected by the paramagnetic impurities for modes laying nearby or in the ESR bandwidth. It has been recently demonstrated that for a mode near 10 GHz there exists a value $P_{0}$ of the injected power specific to each resonator at which the effect of ESR saturation on the mode frequency compensates the impact of the radiation pressure \cite{jap-2014}. Thus the mode sensitivity nulls to the first order, greatly relaxing the specification on the signal power stabilisation.

\subsection{Aging}

Besides pure fluctuations, the properties
of the resonator, or an environmental parameter, may \it drift\rm\  with time. For example, it has been observed that the mechanical stress induced during the resonator's assembly might relax with a long time constant. Should the change in such a stress causes a corresponding change in the wave velocity of the resonator's medium, the oscillator's frequency will drift with time \cite{chang01}. Such phenomena will cause a degradation in the CSO's frequency stability over the longer term. This "aging" is generally characterized in the time domain by an Allan standard deviation increasing proportionally with the integration time $\tau$. As illustrated in the figure \ref{fig:photo-1}, two types of resonator mounting have been tested. Some resonators have a protubating spindle that permits to attach the resonator without perturbing the effective volume in which the electromagnetic wave in confined. Some others are simply fixed a brass screw passing through a 5 mm hole along their axis. In this last case the stress induced by the screw should affect the effective volume. A larger drift is expected for this configuration.    

\begin{table*}[t!!!!]
\centering
\caption{Characteristics of the tested resonators}
\label{tab:X-tal}
\begin{tabular}{lllcll}
\hline
Resonator 	&Year of				&  $\Phi \times H$ 		& Growth  	& Manufacturer & Features\\
Designation	&Devivery 			&(mm $\times$ mm)		& method	 	& 			 &		\\
\hline
\hline	 
HEM-CS 	& 1995 	& 50	x 20			& Heat Exchange &Crystal System & Open cavity experiments \\
HEM-CS	& 2007	& 54 x 30			& Heat Exchange&Crystal System & Optimized 10 GHz CSO, spindle mounting\\
HEM-CS	& 2012	& 54	x 30			&Heat Exchange &Crystal System & Optimized 10 GHz CSO, spindle mounting\\
CZ-PST	&2011	& 54	x 30			& Czochlrasky   &Precision Sapphir Technology & Preliminar samples, screw mounting\\
KY-PST	&2012	& 54 x 30			& Kyropoulos	 &Precision Sapphir Technology &Preliminar samples, screw mounting\\
KY-PST	&2013	& 54 x 30			& Kyropoulos	 &Precision Sapphir Technology &Optimized 10 GHz CSO,screw mounting\\
BAG-CI	&2013	& 30 x 17			& Bagdasarov  &Cristal Innov & Very preliminar sample\\
KY-CI	&2013	& 30 x 17			& Kyropoulos   &Cristal Innov & Very preliminar sample\\

\hline
\end{tabular}
\end{table*}

\section{Sapphire samples} 

\begin{figure}[h!]
	\centering
	\includegraphics[width=\columnwidth]{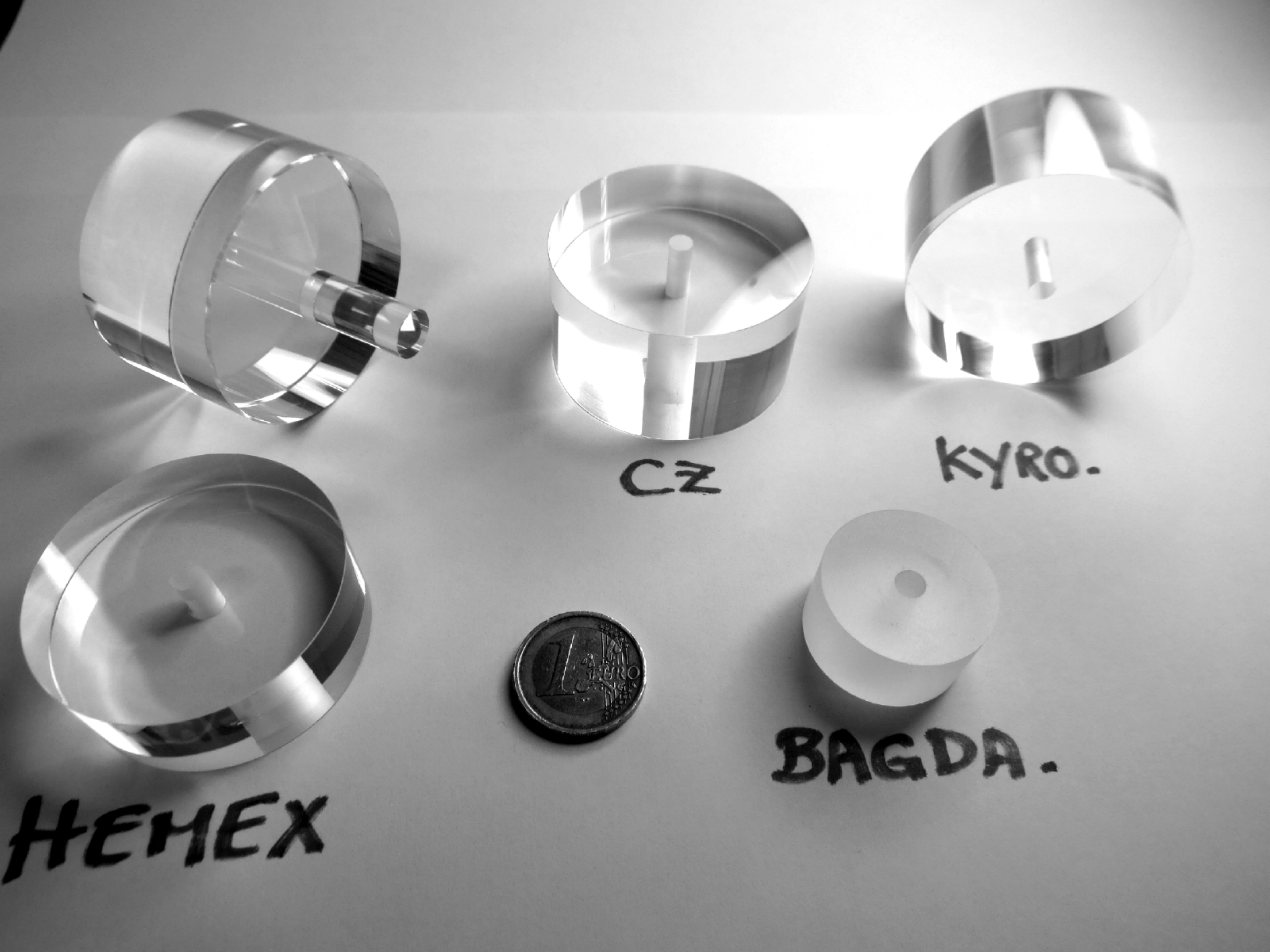}
	\caption{Picture showing different types of sapphire resonators. Up-right, a HEMEX resonator with a spindle. The Bagdasarov crystal is not polished.}
	\label{fig:photo-1}
	\end{figure}

Since our earlier works more than 15 years ago, we have tested different crystals and geometries. Those that have been used for this comparison are listed in the table \ref{tab:X-tal}. Some of them are shown in the figure \ref{fig:photo-1}.

Many techniques are now used for sapphire bulk-crystal growth, see for example \cite{akselrod-2012}. The Heat Exchange Method (HEM) has been commercially developed by the company Crystal Systems (Salem, MA, USA) , now part of GT Advance Technology (USA)\cite{www.crystalsystem}. HEM is known to give high quality sapphire crystals with low dislocation and impurity densities. In the pionner's works leaded at the University of Western Australia (UWA), the superior properties of the HEMEX grad sapphire microwave resonator were demonstrated \cite{luiten-1993-ell}. Since then,  all the CSOs that have been build in different laboratories around the world \cite{luiten94,dick00,uffc04_open_cavity,marra07} incorporate an HEMEX sapphire resonator.
The Kyropoulos technique is also able to produce high quality and large sapphire monocrystals. Nevertheless to our knowledge, its use in a CSO has never been reported. Czochralski crystals are expected to have a higher defect density although Q-factors of the order of 1 billion at 4 K have been already reported \cite{mann-1992-fcs}. We tested Kyropoulos and Czochralski crystals that have been manufactured by Precision Sapphire Technologies (PST- Lithuania)\cite{www.pst}. The last crystals we characterized are very preliminar Kyropoulos and Bagdasarov samples provided by the french technology institute Cristal Innov \cite{www.cristal-innov}. These samples are 30 mm diameter and 17 mm hight and are not polished. The Bagdasarov method is similar to the Horizontal Bridgman technique widely used to grow plates of large surface area. This method produces crystals with a moderate quality.\\

The structural defect density will be affected by the growth rate and the thermal gradient experienced by the crystal during the process. It is thus expected that the Q-factor is dependant on the growth technology and on the skill of the manufacturer. The dopants content is certainly more impacted by the quality of the raw material used. The refractory metal in which the crucible is made can also contaminate the crystal.

\section{First measurements at room temperature}

\begin{figure}[h!!!!]
	\centering
	\includegraphics[width=\columnwidth]{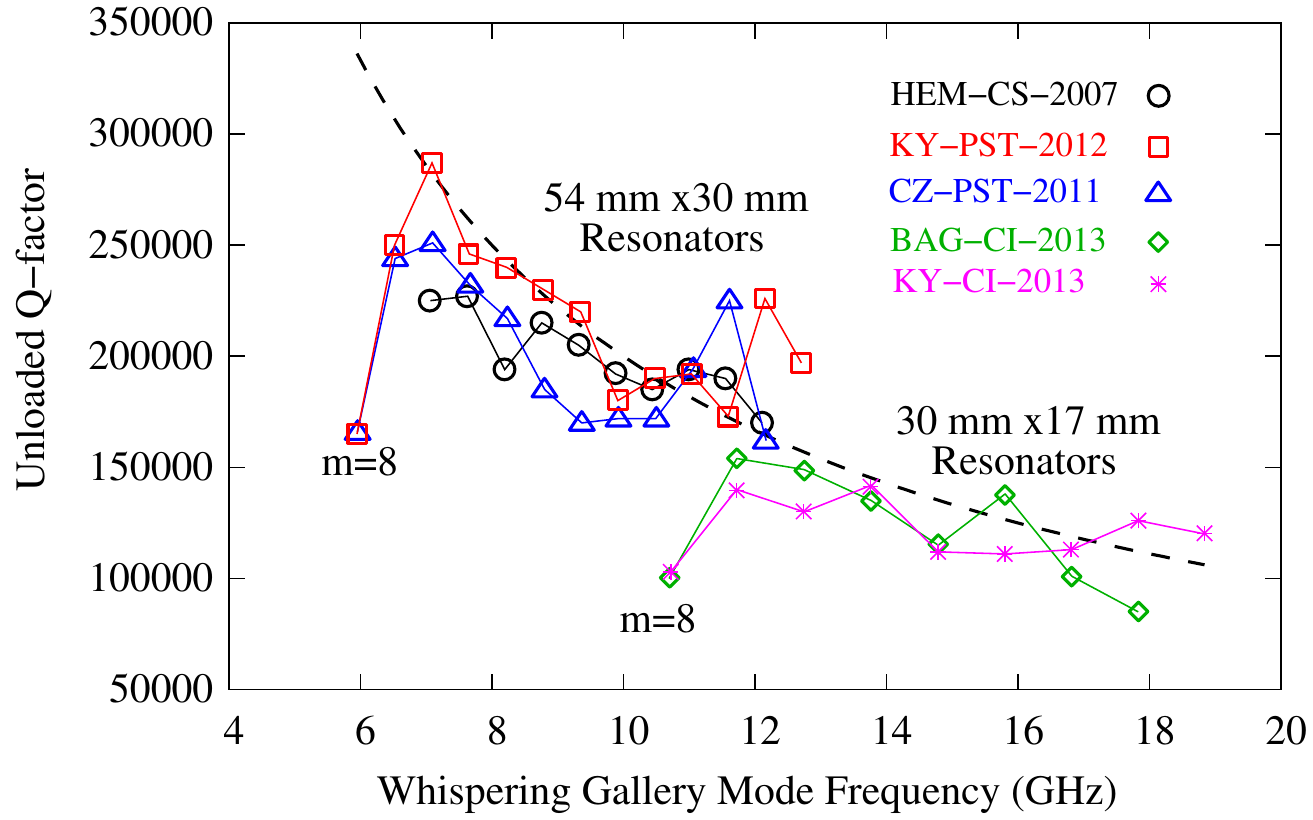}
	\caption{{Room temperature unloaded Q-factor as a function of the mode frequency.}}
	\label{fig:figure-3}
	\end{figure}

The figure \ref{fig:figure-3} shows the unloaded Q-factor measured in an open-cavity configuration for some of the tested crystals.
For all the $54\times30$ mm resonators the mode  $WGH_{15,0,0}$ at 10 GHz presents Q-factor of the order of 200.000, which is a value generaly observed with high quality sapphire pieces \cite{hartnett06-uffc}. As $m$ is sufficiently large, that means the sapphire dielectric loss tangent is $\tan\delta(10 \mathrm{GHz}) =5 \times 10^{-6}$. Starting from this value the dotted line in figure \ref{fig:Q-Tambiante} is the expected Q-factor assuming the sapphire dielectric losses are predominant and proportional to the signal frequency, i.e. $Q_{0}$ varies as $\nu^{-1}$. Modes with low azimutal index show extra radiation losses and thus are not limited by the resonator material. The $30\times17$ mm Cristal Innov samples have Q-high resonances for $\nu \geq 12$ GHz. All the tested crystals present a room temperature unloaded Q-factor limited by the sapphire dielectric losses: no impact of the growth method has been observed at 300K.
\section{Low temperature measurements}
\subsection{Q-factor}
\begin{figure}[ht!]
	\centering
	\includegraphics[width=\columnwidth]{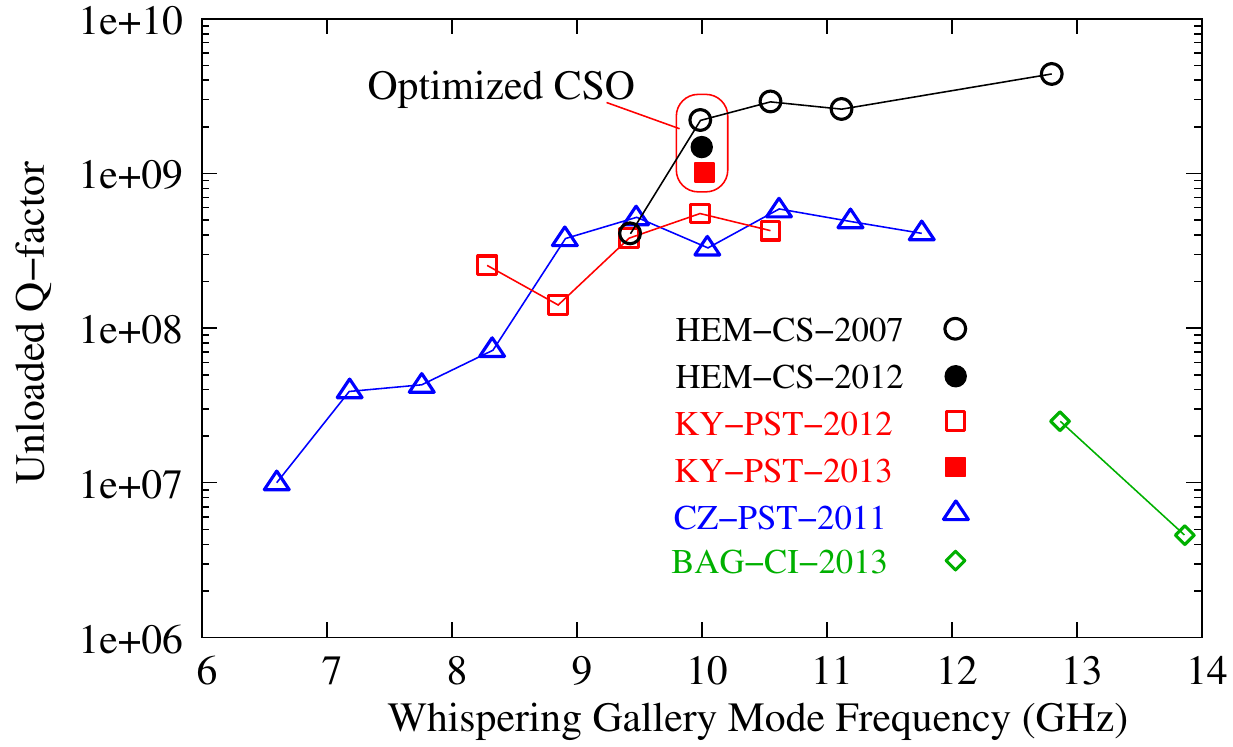}
	\caption{{Whispering gallery mode unloaded Q-factor at 4K.}}
	\label{fig:figure-4}
\end{figure}

The resonators were mounted inside a copper cavity designed for a $54 \times 30$ mm resonator operating on the $WGH_{15,0,0}$ mode at 10 GHz. This assembly is cooled down to 4 K and the resonator unloaded Q-factor is determined from the measurements made with a Network Analyzer (N.A.). The power injected into the resonator is always below $50~ \mu$W to avoid any saturation of the paramagnetic ions. The figure \ref{fig:figure-4} shows the unloaded Q-factors measured at 4 K.

For high azimutal numbers, the unloaded Q-factor is ultimately limited by the resonator material dielectric losses. However, at low temperatures, it can be affected by some other detrimental effects as:
\begin{itemize}
\item Cavity wall losses due to non-optimized geometry
\item Residual contamination of the resonator surface
\item Extra-losses induced by nearby spurious modes 
\item Extra-losses due to the coupling probes
\item Structural resonator defects (dislocations, inclusions, paramagnetic impurities,...)
\end{itemize}
In fact once the operating mode chosen for a given resonator, the cavity has to be designed to limit the spurious modes nearby. Then the couplings have to be adjusted. This fine tuning requires multiple cool down cycles and ideally a careful cleaning at each run. These time consuming procedures have not been strictly followed for all the measurements presented here. Thus this is only for few of them, i.e. those that were used to build an optimized oscillator, that the measured Q-factor can be considered as approching the true material limitation. For the others and following our experience we expect that after careful optimisation and cleaning the Q-factor value can gain up to $20\%$ of its preliminary value.\\

The HEMEX grad resonator shows superior performances attaining 2 billions for the operating mode at 10 GHz. The Czochralski  and Kyropoulos resonator from PST can achieve better than $5\times 10^{8}$ unloaded Q-factor, and thus are suitable to meet the short term frequency stability of $1\times 10^{-15}$. The cavity geometry is not adapted for the resonators provided by Cristal Innov. For these crystals the spurious mode density was too high to find high-order whispering gallery modes. We report here only the modes $WGH_{10,0,0}$ (12.87 GHz) and $WGH_{11,0,0}$ ($13.86$ GHz) we observed in the Bagdasarov resonator, which obviously presents detrimental radiation losses. 
Further experimental investigations have to be conducted on these crystal to determine the actual material losses.

\subsection{Turnover temperature}
\begin{figure}[h]
\includegraphics[width=\columnwidth]{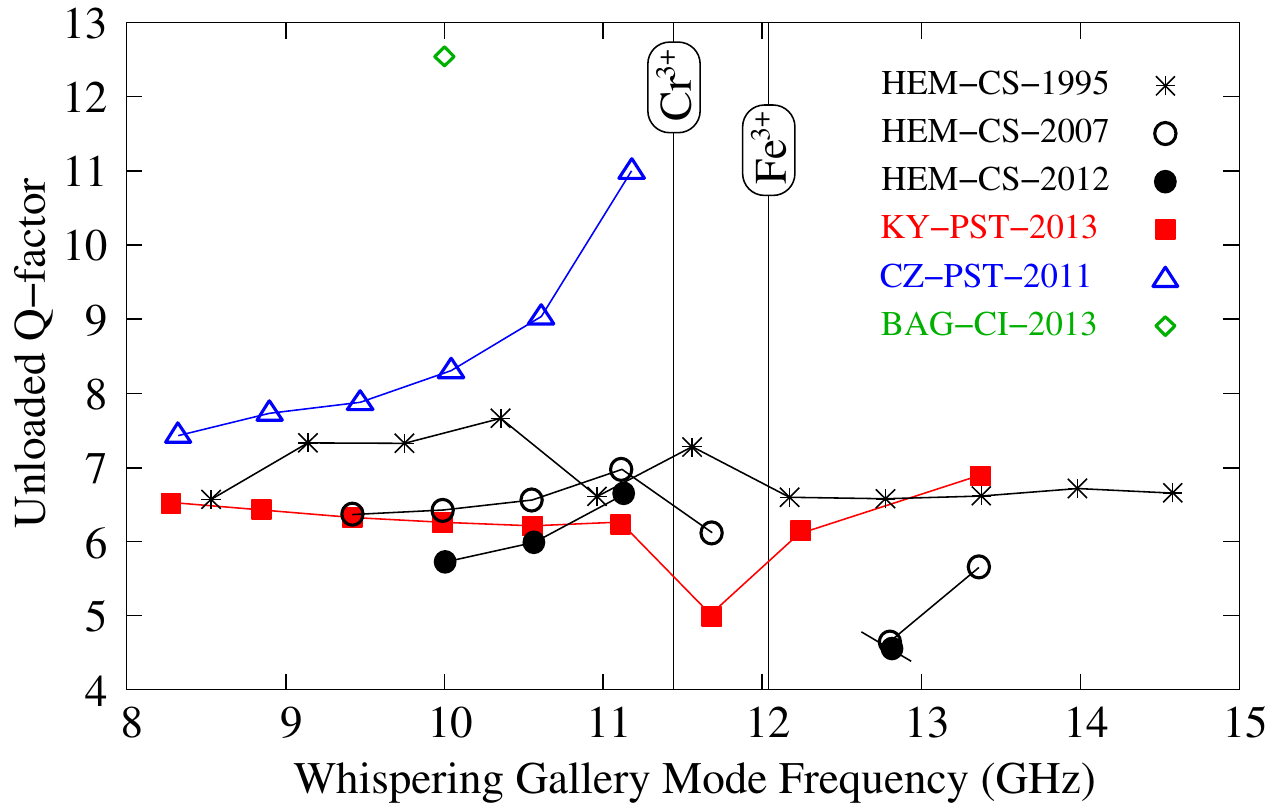}
\caption{Turnover temperature as a function of the mode frequency.}
\label{fig:figure-5}
\end{figure}

To determine the turnover temperature, the resonator temperature set-point is increased step by step. At each step and after waiting for the system stabilisation the modes resonance frequencies are measured with the N.A. referenced to a Hydrogen Maser to ensure long term stability. Eventually for each mode presenting a turnover the collected data are fitted with a second order polynome to calculate $T_{0}$. Figure \ref{fig:figure-5} summarizes the results for all the crystals we tested. It gives the turnover temperature as a function of the mode frequency. \\

\begin{figure}[h!]
	\centering
	\includegraphics[width=\columnwidth]{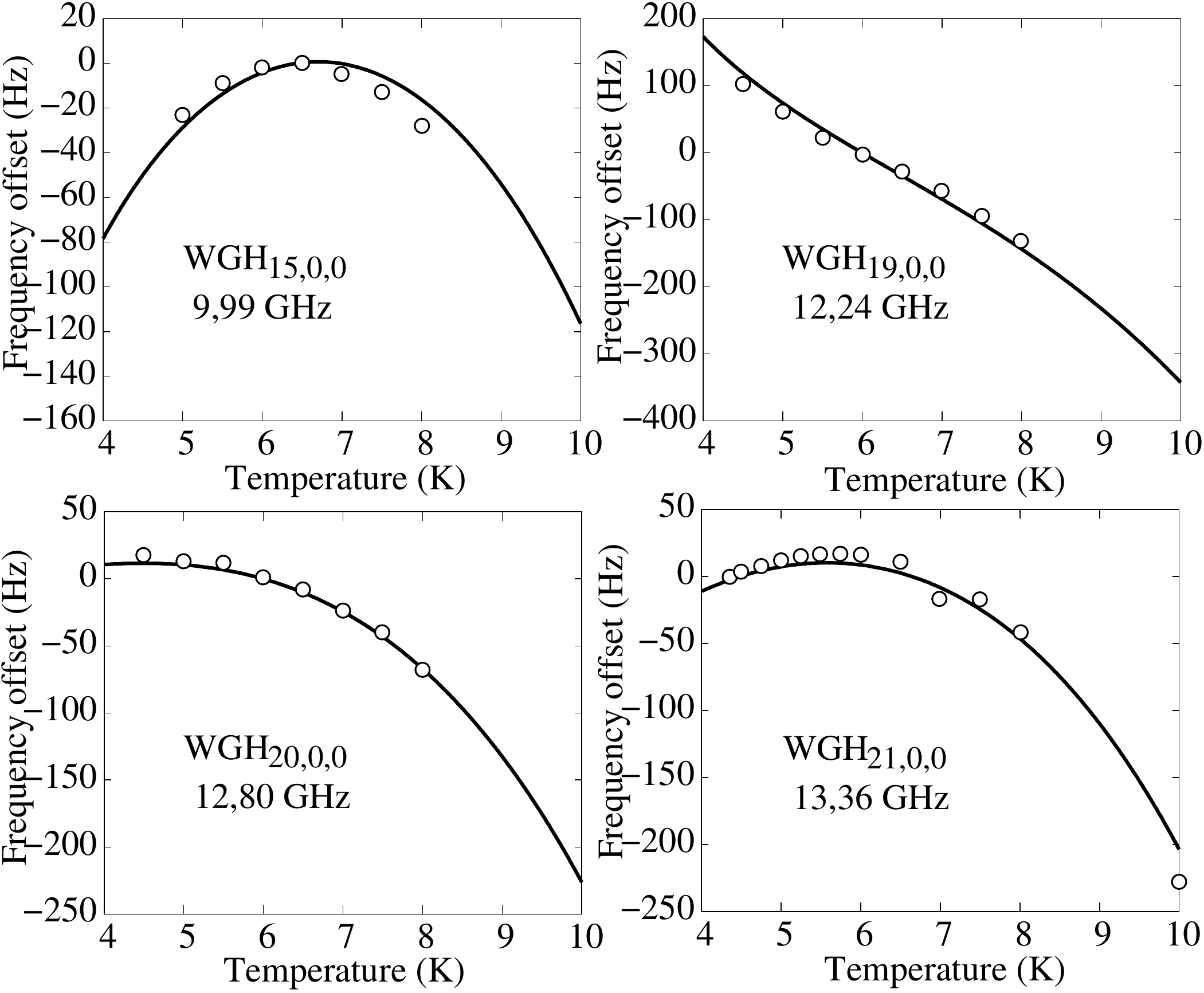}
	\caption{HEM-CS-2007 resonator frequency-to-temperature $\nu(T)$ for the modes $WGH$ $m=15,19,20$ and $21$. Open circles: experimental data. Bold lines: simulated thermal behaviors assuming  150 ppb of  Mo$^{3+}$, 3 ppb of Cr $^{3+}$  and 3 ppb of  Fe$^{3+}$. The origin of the frequency offset has been arbitrary chosen.}
	\label{fig:figure-6}
	\end{figure}

Figure \ref{fig:figure-6} presents for the HEM-CS-2007 resonator the experimental frequency-to-temperature $\nu(T)$ dependances for modes placed on each sides of the Cr$^{3+}$ and Fe$^{3+}$  ESR frequencies. The simulated behaviors are reported in bold lines.

Equation \ref{equ:f-T-sens} was used to compute $\nu(T)$, the susceptibility $\chi'(\nu)$ being the summation of the contribution of all ion species: Cr$^{3+}$, Fe$^{3+}$ and Mo$^{3+}$. 
As explained in \cite{jap-2014}, the individual dc-susceptibilities $\chi_{0}$ have been evaluated with the Van Vleck equation \cite{vanvleck78}. Indeed the Curie law has been derived for a free system of spin $S$, which consists in $2S+1$ levels equally spaced. In a real crystal, the ground state is splitted by the crystal field in multiple degenerated Kramer’s doublets separated by a Zero Field Splitting and strictly speaking the dc-susceptibility should be calculated by using the Van Vleck equation. Nevertheless the Curie deserves to be introduced as it makes explicite the $1/T$ dependance of the susceptibility. Moreover the difference in the $\chi_{0}$ values calculated with the two models is less than $20\%$ in the temperature range of interest. The modes thermal behavior have been correctly simulated by assuming the densities given in table \ref{tab:3} for each expected dopant. \\

In old samples provided in the 90s (HEM-CS-1995), all whispering gallery modes in a large frequency range present a turnover temperature almost independent of the mode order (see figure \ref{fig:comparaison-Tinv}). Luiten \cite{luiten96} demonstrated that it is due to the predominance of the Mo$^{3+}$ ion, whose ESR frequency is 165 GHz. The spread in turnover temperatures observed for low frequency modes ($\nu <12$ GHz) could result from Cr$^{3+}$ or/and Fe$^{3+}$ residuals. The concentration of these residuals should be very low as the turnover temperature imposed by the Mo$^{3+}$ ions is not greatly affected.\\

In most recent HEMEX crystals, it appears that the relative concentrations of Cr$^{3+}$ and Fe$^{3+}$ are higher. Starting from the lowest frequencies, the turnover temperature increases as the mode approaches the Cr$^{3+}$ ESR. HEM-CS-2007 resonators keep a turnover temperature between $\nu_{Cr}=11.44$ GHz and $\nu_{Fe}=12.04$ GHz but not for the $WGH_{19,0,0}$ at 12.24 GHz just above the Fe$^{3+}$ ESR. The $WGH_{20,0,0}$ mode at 12.80 GHz recovers a turning point but at a lower temperature, i.e. 4.6 K. For the HEM-CS-2007 resonators the impact of the  Cr$^{3+}$ and Fe$^{3+}$ ESR are clearly visible on the figure \ref{fig:comparaison-Tinv}: $T_{0}$ follows a dispersive like curve resulting from the summation of the two dispersive lonrentztians centered on $\nu_{Cr}$ and $\nu_{Fe}$. In HEM-CS-2012, the Cr$^{3+}$ density seems to be higher as no turnover is observed between $\nu_{Cr}$ and $\nu_{Fe}$.\\

For the Kyropoulos resonator (KY-PST-2013) all the modes between 8 and 13 GHz present a turnover temperature near 6 K apart the $WGH_{18,0,0}$ (11.72 GHz) mode, for which $T_{0}=5.4$ K. The behavior of this resonator is very close to those of the HEMEX grad resonators with certainly a lower Fe$^{3+}$ density.\\

The Czochralski crystal shows turnover temperatures only for the WG modes below the Cr$^{3+}$ ESR, indicating that this ion is the predominant paramagnetic impurity.\\

The turnover temperature as well as the shape of the thermal behavior of each mode depend on the dopants densities. By assuming the crystals contain only Cr$^{3+}$, Fe$^{3+}$ and Mo$^{3+}$, we have been able to simulate correctly the thermal behavior of each modes as demonstrated in figure \ref{fig:Tinv-alizee-15-19-20-21-v2} for the HEM-CS-2007 crystal.  The effective paramagnetic ion densities that permit to simulate correctly the thermal behavior of each crystal are given in table \ref{tab:3}. 

\begin{table}[h!!!!!]
\centering
\caption{ESR of the Iron-group paramagnetic ions that can be found in high-purity sapphire crystals.}
\label{tab:3}
\begin{tabular}{lcccc}
\hline
Ion	&$S$	&$\nu_{j}(GHz)$	&$\tau_{2}$ (ns)	&$T_{0}(K)$ \\
	&		&        			&				&$1$ppm/10 GHz			\\	
\hline
\hline
Cr$^{3+}$  	&3/2		& 11.4~~	 	& 7	&		12.6	\\
Fe$^{3+}$  	& 5/2	& 12.0~~ 		& 20 &		 13.9\\
Mo$^{3+}$  	& 3/2	& 165.0~~		& 12	&		 8.4\\
\hline
\end{tabular}
\end{table}

\section{CSO frequency stability}
\begin{figure}[h]
\includegraphics[width=0.9\columnwidth]{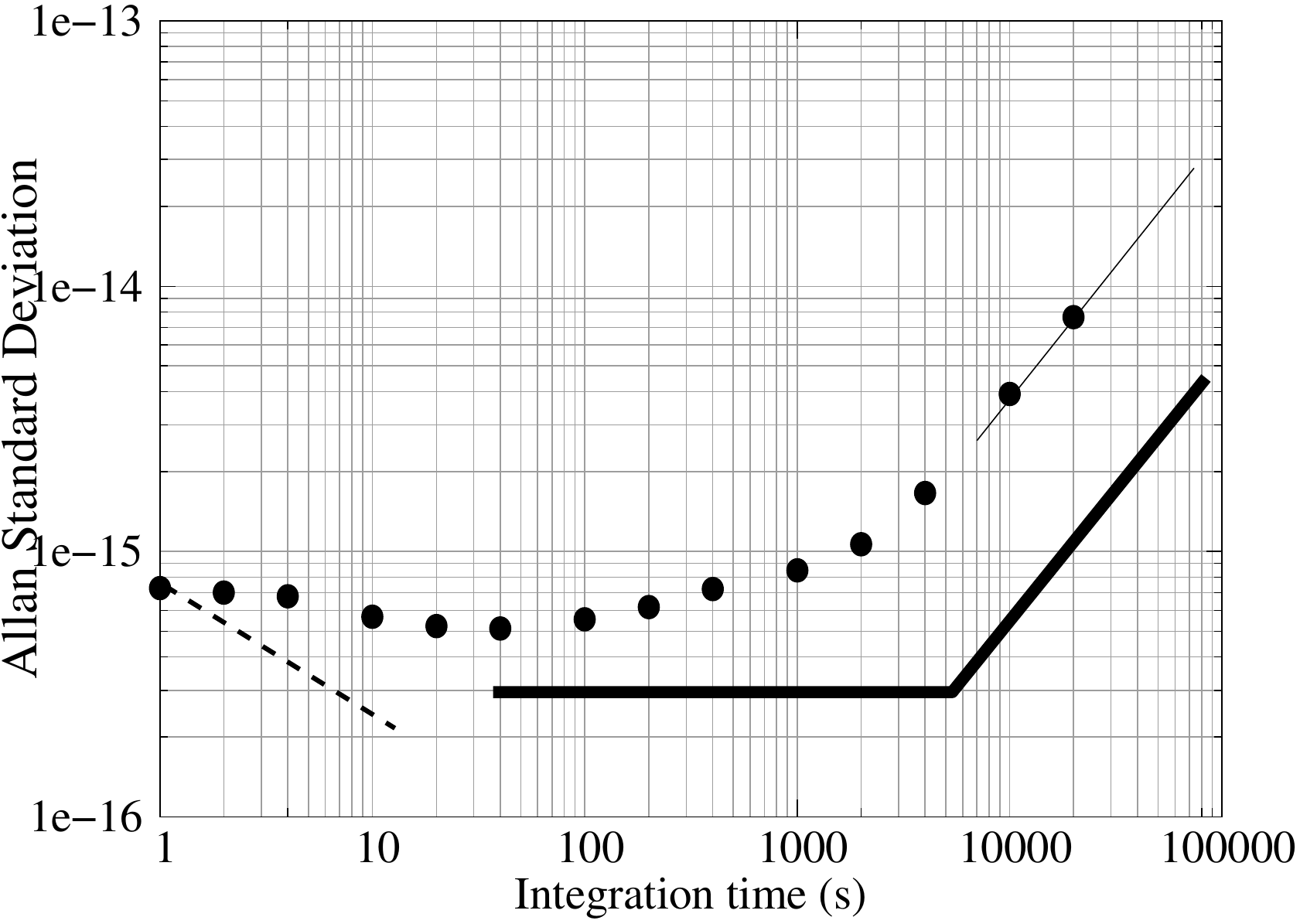}
 \caption{Allan standard deviation calculated from the beatnote between two 10 GHz CSOs. Black bullets: one of the CSOs is equiped with a Kyropoulos crystal from PST, the other one incorporates an HEMEX grad sapphire crystal. Bold line: the best stability we obtained with a HEMEX resonator \cite{jap-2014}. Dashed line: instrument noise floor.}
 \label{fig:figure-7}
\end{figure}

Until now, due to its superior properties the HEMEX crystal was always choosen in the realization of an ultrastable microwave cryogenic oscillator. The previous results show that the Kyropoulos crystal can also constitute a good material for a cryogenic resonator. We thus incorporated the KY-PST-2013 resonator in one of our CSO. Its fractionnal frequency stability has been determined by beating its output with another CSO equiped with a HEMEX grad resonator. The resulting Allan standard deviation is given in the figure \ref{fig:figure-7}.

The apparent flicker floor obtained with the Kyropoulos resonator is $\sim 5\times 10^{-16}$ whereas it was $3\times 10^{-16}$ with an optimized HEMEX CSO \cite{jap-2014}. Such a difference can be attributed to the lower Q-factor of the Kyropoulos resonator. As expected the long term drift of Kyropoulos CSO is higher than those observed with the HEMEX CSO. As ralready mentionned, the Kyropoulos crystal is simply maintained by a brass screw passing through the hole along its axis. The resulting stress induced into the effective resonator volume relaxes with time, leading to a drift 10 times larger than the drift observed when the resonator is equiped with a spindle. 

\section{Conclusion}

We tested sapphire crystals produced with different growth methods in order to determine which type is suitable to build an ultra-stable cryogenic oscillator.
At room temperature the Q-factor of all the tested crystal is limited by the sapphire loss tangent, i.e $\tan\delta =5 \times 10^{-6}$ at 10 GHz. At low temperatures the HEMEX and the Kyropoulos resonators both present a Q-factor higher than $5\times 10^{8}$ and a turnover temperature compatible with an operation in a cryocooler. These resonators have been used as the frequency determining element of a cryogenic oscillator and a fractional frequecy stability  better than $1\times 10^{-15}$ at short term has been obtained in both cases. Furthermore the analysis of the modes thermal behavior between 4K and 12K allows us to approximatly determine the impurities contents of each crystals.

\section*{Acknowledgment}

The work has been realized in the frame of the ANR projects: Equipex Oscillator-Imp and Emergence ULISS-2G. The authors would like to thank the Council of the R\'egion de Franche-Comt\'e for its support to the Projets d'Investissements d'Avenir and the FEDER (Fonds Europ\'een de D\'eveloppement Economique et R\'egional) for funding one CSO.

\end{document}